\documentclass[twocolumn,aps,prl,english,twocolumn,epsfig,groupedaddress]{revtex4}
\usepackage{graphicx}
\usepackage[latin1]{inputenc}

\begin{document}
\title{\bf Model for vacancy-induced d$\sup 0$ferromagnetism in oxide compounds.}

\author{ Georges ~Bouzerar$^{1}$\footnote[6]{email: georges.bouzerar@grenoble.cnrs.fr} and Timothy ~Ziman$^{2}$\footnote[5]{and CNRS, email: ziman@ill.fr} 
}

\affiliation{ 
$^{1}$Laboratoire Louis N\'eel 25 avenue des Martyrs, CNRS, B.P. 166 38042 Grenoble Cedex 09
France.\\   
$^{2}$Institut Laue Langevin B.P. 156 38042 Grenoble 
France.\\
}            
\date{\today}

\begin{abstract}
\parbox{14cm}{\rm}
\medskip
We propose a model with few parameters for vacancy-induced ferromagnetism
based on a correlated model for oxygen orbitals with random
potentials  representing cation vacancies. For certain potentials, moments appear on oxygen sites
near defects. Treating the randomness exactly, we calculate the magnetic couplings between moments, the Curie temperature and spin and charge densities as a function of the potential, the density of vacancies, and
correlation strength. For physically reasonable parameters this predicts Curie temperatures well above room temperature for small concentrations of vacancies. We discuss our results in relation to questions of stability and reproducibility raised in experiments. To circumvent the difficulties of controlling intrinsic defects, we propose specific non-magnetic host doping that could be, for example, substituted for cations in HfO$_2$ or ZrO$_2$.
\end{abstract} 

\maketitle

\section{}

A number of exciting, but puzzling, effects when ferromagnetism is seen in
unexpected places, are apparently unrelated to traditional transition metal magnetism: in  oxides such as  HfO$_2$, ZrO$_2$\cite{CoeyNature}, CaO, ZnO,  and related materials such as hexaborides\cite{Boride}
(CaB$_6$) and even irradiated graphite\cite{Graphite1,Graphite2}. 
Thin films of materials which in bulk 
have neither magnetic moments nor magnetic order, 
may be ferromagnetic well above room temperature\cite{Coey2,CoeyHfO2}.
 HfO$_2$ or ZrO$_2$, are both wide-band insulating material
with high dielectric constant; the possibility of making them ferromagnetic
could widen their possible application in the field of spintronics.
There are  analogies to  the discovery  \cite{Ohno} that the doping by a small amount of magnetic impurities could lead to relatively high Curie temperature in III-V semiconductors. There are, apparently, however too few magnetic impurities in the ZrO$_2$ films, for example, to explain the  magnetism and the cation (nominally Zr$^{4+}$) should be in a non-magnetic d$^0$ configuration.
The term "$d^{0}$" ferromagnetism was coined to describe this
general phenomenon. There are suggestions that  the ferromagnetism  is related to  intrinsic cation or anion vacancies, and that this may be a path to  new ferromagnets\cite{Stoneham,Sawatzky}.Band structure calculations\cite{Monnier,Sanvito} supported the idea that vacancies can induce local moments on neighboring atoms. What is lacking, especially as the experimental results are still unstable, is a quantitative theory of 
long-range ferromagnetic order of any such moments. This can define the important parameters, for example the nature and concentrations of vacancies and their effective doping. The aim of the present letter is to provide a model for vacancy induced ferromagnetism explaining Curie temperatures at room temperatures. This can also help understand instabilities, and propose new materials where controlled substitutions could replace the vacancies. Point defects have been suggested several times as a mechanism for localized moments\cite{Stoneham,Sawatzky,Zunger} and ferromagnetism in oxides.
While oxygen vacancies were originally suggested as the source
of magnetism in HfO$_2$ \cite{CoeyHfO2}, density functional calculations including various  defects \cite{Sanvito} found no evidence for moments around such defects. A large magnetic moment was, however, found in partially depleted oxygen orbitals around Hf vacancies. Couplings were estimated between two defects confined within a supercell. This is not sufficient to estimate properly 
the Curie temperature as a function of vacancy concentration which require couplings at all distances. We propose a theory where dependence on concentration and also the effective doping of different defects can be studied. We start from the idea of vacancy-induced moments
and construct a model, with only few free parameters, in which we can estimate not only the temperature dependence of local moments, but also the couplings required to maintain long-range ferromagnetic order at the low concentrations relevant to the experiments. We can also calculate the thermodynamics and dynamical spin correlations for future systematic experimental investigations.
We formulate the problem for the  cases of 
cation  vacancies in  oxides such as HfO$_2$, ZrO$_2$ and CaO,
but the aim is to explore what may be a more geneal
phenomenon, in which a vacancy or substitutional defect creates an {\it extended} magnetic moment on neighboring atomsWhile there are similarities to the theory of magnetism in diluted magnetic semiconductors
but there the substitutional impurities immediately provide localized magnetic moments which interact via the itinerant carriers of the (doped) host. 
 Here \cite{Sawatzky} the moment is induced
on several non-magnetic oxygen atoms of the original host
and their is no clear separation between the holes forming the 
moment and those mediating the interactions.
\par
We describe this  via a single correlated band
of oxygen orbitals, with random potentials representing the influence
of the randomly introduced vacancies on neighboring cation
orbitals. The oxygen atoms are on a regular lattice, taken to be cubic
for simplicity. In the pure host, the parameters
can be those calculated for the filled oxygen-dominated
bands. The cations in the pure material are assumed to be at the center of 
each elementary cube of the lattice. A concentration $x$ of cation vacancies
is modeled  by choosing randomly the positions of
cube centers. Thus a ``defect'' does not alter the regular geometry of the oxygen lattice but introduces attractive potentials on the 8 neighboring oxygen orbitals of the cube center chosen. The usual Hubbard
Hamiltonian on a regular lattice has  an extra term which is a random, but {\it locally
correlated}, potential: 
\begin{eqnarray}
 {\cal H}_0= - \sum_{{i},{i}^{\prime},\sigma}  t_{{i},{i}^{\prime}} \left({\bf c}_{{i}}^{\dag\ \sigma}\cdot{\bf c}_{{i}^\prime}^{\sigma} +{\bf c}_{{i}}^{\sigma}\cdot{\bf c}_{{i}^\prime}^{\dag\ \sigma}
 \right) \nonumber \\
+ \sum_{{i}} U  {\bf n}_{{i}}^{\uparrow} {\bf n}_{{i}}^{\downarrow}
- V_{{i}} \left({\bf n}_{{i}}^{\uparrow}+  {\bf n}_{{i}}^{\downarrow}\right) 
\nonumber
\end{eqnarray}
$c^{\dagger}$ ($c$) operators create (destroy) electrons in the oxygen orbitals.
$V_i$ are attractive on sites adjacent to the 
vacancies, corresponding to attraction of the holes. 
Nearest neighbor hopping $t_{i,j}=t$ everywhere except between pairs of sites
neighboring the same vacancy for which  $t_{i,j}=t^{\prime}$. 
Defects in Hf/ZrO$_2$ have two inequivalent sets of neighboring
oxygens (O1 and O2)\cite{Sanvito}.
To ease comparison with Local Density Approximation density of states we took $V_{i}=V$ on four neighbors and $V_{i}=\frac{V}{3}$ on the other four. $V$ is a free parameter.
The hole density is taken to be a variable  $n_h$ . In the simplest picture ({\it only} cation vacancies) $n_h$ is  $4 x$ for HfO$_2$ and ZrO$_2$ (cation 4+) and  $2x$ for CaO or ZnO (cation 2+). In numerics we quote $U$ and $V$ in terms of the host bandwidth $W=12t$.
For ZrO$_2$ we take $W= 7~eV$ from the calculated bandwidth
of the p band just below the Fermi level \cite{Vanderbilt}
and $U\sim 3$ eV as estimated for the Hubbard parameter for oxygen p orbitals U$_{pp}$ in copper
oxides \cite{Martin}.   
\par
\begin{figure}[tbp]
\includegraphics[width=8.cm,angle=-90]{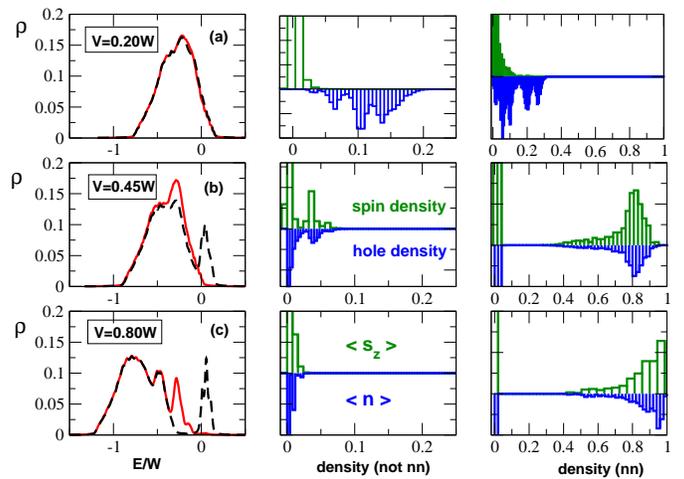}
\caption{(Color online)
Density of states (left) and distributions 
of charge and magnetic moments (middle and right column). 
The right column shows the densities projected onto oxygen neighboring
vacancies (nn); the middle column other oxygens (not nn).
The majority and minority bands(left column) are continuous and dashed
respectively. Energies are in units of bandwidth measured from the Fermi level,
the moments in units of $\mu_B$.}
\label{dos}
\end{figure}

We treat the correlated disorder {\it exactly} as
the unrestricted Hartree-Fock (UHF) approximation is applied
in real space for each configuration of disorder, sampled over many
configurations. Comparison to exact results indicate that UHF works well for ground state properties in random systems\cite{BouzerarCurrents,Grabert}.All calculations presented were for sufficiently large systems (typically $\sim 16^3$ sites) that finite size effects are negligible.
In Fig.\ref{dos} we show the single particle density of states and the distributions of both hole density and induced local moments for fixed $U=0.4~W, n_h=0.12, x=4\%$ and three values of $V=0.2$, 0.45, 0.8 W. The Fermi energy is at zero. For small V there are no local magnetic moments. As V increases, moments appear  ($V\ge 0.28~W$) as the charge and spin distributions develop structures localised around the vacancies (right column). Later ($V\approx 0.45~W$), a well-defined impurity band for the minority band splits from the valence band. The first peak in the distributions (right column), at low densities, corresponds to sites with $V_{i}=V/3$ and the second broad peak to $V_{i}=V$. As V further increases, we also observe a splitting of the density of states in the majority band and holes concentrated around the vacancies, providing saturated localized moments. Note that the total moment on each cube centered on a vacancy is almost saturated ($\sim 3 \mu_B$) for $V \ge 0.4~W$.  
We now estimate the magnetic couplings at all distances between the different moments,by UHF calculation of the carrier Green functions $G_{{i} {i}^{\prime}}^{\sigma}$.
We will subsequently make a  self-consistent local random phase approximation (SC-LRPA) on the effective random Heisenberg model\cite{Bouzerar2}, which was successful for diluted magnetic semiconductors. 
Each  spin of the Heisenberg model will represent the {\it total} spin around one vacancy.
Calculation of the exchanges is made by an extension of the method of Lichtenstein {\it et al.}\cite{Lichtenstein}. They determined the exchange between any pair of (local) moments by calculating
the change in energy when fields are applied to the sites of the two moments.
Here, the fields are applied to two differing directions, the first to {\it all } the $N_S=8$ neighbors ${i}$ of one defect ${a}$; the second to the neighbors ${j}$ of the second ${b}$. Thus we calculate the exchange energies by summing the contribution of $N_S\times N_S$ pairs of terms
${\cal J}_{{a},{  b}}= \sum_{{  i} \epsilon {a},{  j} \epsilon {  b}}{\cal J}_{{  i},{  j}}
$
where 
\begin{eqnarray}
{\cal J}_{{  i},{  j}}= -{\frac {1} {\pi}} \Im \int_{\infty}^{E_F} \Sigma_{{  i}}(\omega)G_{{  i},{  j}}^{\uparrow}(\omega) \Sigma_{{  j}} (\omega)G_{{  j},{  i}}^{\downarrow}(\omega) d\omega\nonumber
\end{eqnarray}
In UHF the local potential $\Sigma_{  i}= U(n_{  i}^{\uparrow}-n_{  i}^{\downarrow})$.

The dilute Heisenberg model is defined with spins centered on the position of the randomly
placed vacancies and interacting by the ${\cal J}_{{a},{b}}$.
${\cal H}= -\sum_{a,b}{\cal J}_{a,b}S_a \cdot S_b$ with $S_{a}$  the total spin around each
vacancy. $S_a\sim 3\mu_B$ are large enough to be treated classically, allowing
application of the local force theorem. 
\begin{figure}[tbp]
\includegraphics[width=7cm,angle=-90]{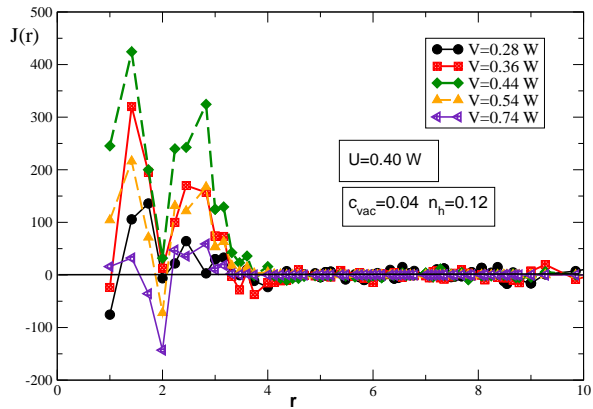}
\caption{(Color online) Magnetic couplings (in Kelvin) as a function
of separation of vacancies, in units of the cubic lattice spacing,
for different values of the potential V. In each case $U= 0.4W$, $x=0.04$ and $n_h =  0.12$
as in Fig.~\ref{dos}. }
\label{couplingspace}
\end{figure}
In Fig.\ref{couplingspace} we show the calculated couplings averaged
over different impurity configurations. We vary the potential $V$ and fix $x= 4\%$ and $n_h/x=3$. This last choice (rather than 4) will be made clearer by the figures following.
For small V the couplings oscillate with distance, but with antiferromagnetic nearest neighbor coupling.
As V increases the couplings become more ferromagnetic but with further increase some become antiferromagnetic again. Thus from Fig.\ref{couplingspace} there is a range of values V where the couplings, while fluctuating, are all ferromagnetic ($J(r) \ge 0$). This is associated
in Figure \ref{dos} (b) with the incipient development of a
visible impurity band just at the band edge. This ferromagnetic bias in the ``RKKY-like'' oscillations corresponds to the resonant form of the impurity band\cite{RichardBouzerar}. 
\par
We  now treat the thermodynamics 
within SC-LRPA\cite{Bouzerar2}.
We recall that randomness is
treated {\it exactly}, essential at the low concentrations relevant here. 
\begin{figure}[tbp]
\includegraphics[width=7cm,angle=-90]{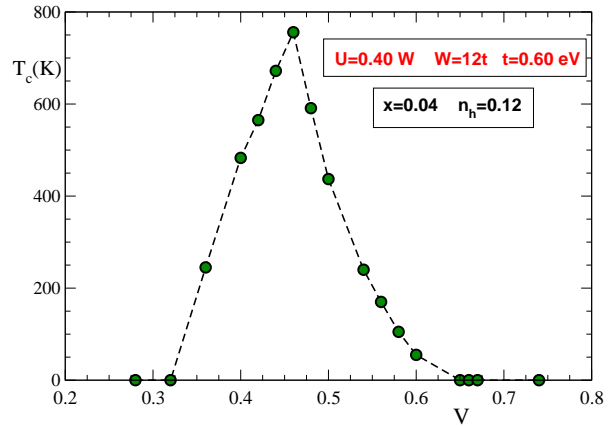}
\caption{(Color online) Curie temperature as a function of the potential V (in units of W). Other parameters as in Fig. \ref{couplingspace}.}
\label{TcV}
\end{figure}
In Fig.~\ref{TcV} we show the Curie temperatures (T$_C$) calculated from the exchanges as a function of $V$. It is seen that there is a well-defined region where T$_C$
is above room temperature. Comparing to Figs.~\ref{dos} and \ref{couplingspace}, we see that for $V$  either too small or too large, T$_C$  vanishes because the couplings are frustrated by
RKKY-type oscillations (small V) or superexchange (large V). 
As seen in Fig~\ref{couplingspace}, near the optimal value (V=0.45W) the couplings are largest  and do not change sign. Remark that at that value (see Fig~\ref{dos}(b)) the
impurity band is very similar to that of doped Ga(Mn)As\cite{Josef}. 
Optimal values of V have high T$_C$ for the same reason, the position of the impurity
band, that Ga(Mn)As has a larger T$_C$ than Ga(Mn)N and In(Mn)As which
resemble more cases (c) and (a) respectively\cite{Bouzerar2}. 
\begin{figure}[tbp]
\includegraphics[width=7cm,angle=-90]{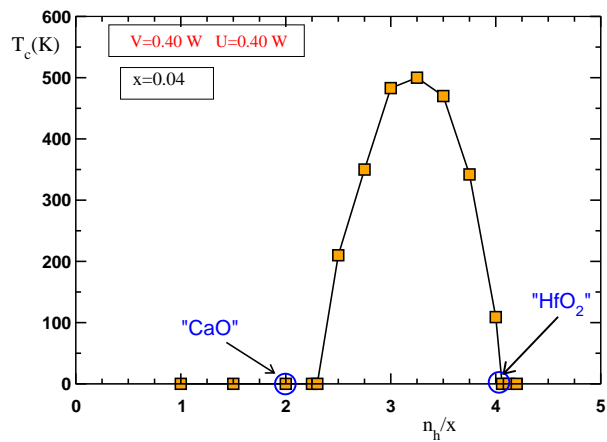}
\caption{(Color online) Curie temperature 
for different carrier densities for $4\%$ defects. $V=0.4W$, $U=0.4W$ are fixed.} 
\label{Tcdoping}
\end{figure}
\par
In Fig.~\ref{Tcdoping} we fix the density of defects and $V$ near the 
optimal value of Fig.~\ref{TcV} and vary the number
of holes per vacancy. Below and above critical values
of $n_h/x$ there is no ferromagnetism. There is a window
of concentrations where T$_C$ becomes very large. The reason for instability outside the window
is for low carrier densities, dominant antiferromagnetic superexchanges and for high densities, RKKY-like oscillations. Note that for the points where $T_C = 0$ our calculation predicts that are still local moments, in agreement, for example, with {\it ab-initio} calculations\cite{Sawatzky}
for defects in CaO, but no long-range ferromagnetic order. 
Interestingly, the formal charges of vacancies in HfO$_{2}$ or CaO are 
near the edges of stability of ferromagnetism for the parameters chosen. This may suggest an explanation for the extreme sensitivity of current results to sample history.
Note that a more direct comparison with experiments would require a precise knowledge of the
parameters and the concentrations of all defects including especially Oxygen vacancies.
Oxygen vacancies, would {\it decrease} the effective hole density and could stabilize T$_C$ for HfO$_2$. The figure also suggests that a promising avenue for stabilizing or increasing
the Curie temperature experimentally may be to use non-magnetic substitution of Hf(Zr) by  Li, Na, K, Rb or Cs .. rather than Hf(Zr) vacancies which are difficult to control.
\par
Our results are based on a Hartree-Fock approximation that tends to overestimate the tendency
to ferromagnetism in some homogeneous itinerant models. Nonetheless Kanamori\cite{Kanamori}, Tasaki\cite{Tasaki} and others\cite{Nagaoka,Mielke,Ohkawa} have shown
that in ``flat band'' models in which there is a peak in the density of states with a relatively flat background, ferromagnetism is indeed possible. Because of the vacancies, our model resembles a flat band model. We remark that degeneracy of orbitals \cite{Sawatzky} is not required, as
the moments are stabilized by the molecular fields of other moments. Note that screening effects which reduce the effective U by the t matrix\cite{Kanamori} are important when $U \gg W$ which is not the case here. In addition, we have verified tat the results vary little with  $t'/t$ and $U$ in broad ranges, $ 0.2 \le U/W \le 0.6$ and $0.1 \le t'/t \le 0.6 $. In contrast $T_{C}$ is very sensitive to both $V$ and $n_{h}/x$ which are the most relevant parameters. We have also seen that below typically $1\%$ no ferromagnetism is possible.In ref.\cite{Zunger} the equilibrium density of cationic vacancies was estimated as $0.03\%$. Thus higher non equlilibrium concentrations are very likely needed, as are quite possible in films. However a more promising avenue for ``d$^0$' ferromagnetism would be the subtitutional approach.

To conclude, we present a model which 
catches the essential ingredients for vacancy- or substitutional-induced ferromagnetism of the oxygen holes. Our calculations shows that Curie temperatures  above room temperature are feasible for a few per cent of vacancies or substitutions. The sensitivity of ``d$^0$''samples may be explained by the  proximity of the doping to the stability boundary. The underlying mechanism is close to that of diluted III-V semiconductors: 
enhanced ferromagnetic couplings between resonant impurity levels at parameters
just at the point where the impurity states split off from the 
valence band. It would be interesting to measure the profile of the density of states (e.g. by photoemission) to see whether it does indeed correspond to the proposed picture (see Fig. \ref{dos} center row). As Curie temperatures are non-monotonic in both the strength of the potential and the doping, accurate characterization may be needed in order to find useful ferromagnetic materials. {\it Ab-initio} results  for the effective Hamiltonian could
help refine the choice of compound. Our calculation, Figure \ref{Tcdoping}, suggests that for HfO$_2$ (ZrO$_2$), site substitution of the Hf(Zr) by  elements  differing by 3 in formal charge, i.e. Group 1A of the periodic table would be a promising direction. This may be attractive since it can be realized in bulk as well as films.


\begin{thebibliography}{1}
\bibitem{CoeyNature} M. Venkatesan, C. B. Fitzgerald, and J. M.D. Coey, Nature (London) {\bf 430}, 630 (2004). 
\bibitem{Boride} D. P. Young {\it et al.}, Nature (London) 397, 412 (1999)
\bibitem{Graphite1} T. L. Makarova {\it et al}., Nature (London) {\bf 413}, 716 (2001).
\bibitem{Graphite2} P. Esquinazi {\it et al}, Phys. Rev. B {\bf 66}, 024429 (2002). 
\bibitem{Coey2}  M. Venkatesan, C. B. Fitzgerald, J. G. Lunney, and J. M. D. Coey, Phys. Rev. Lett., {\bf 93}, 177206 (2004).
\bibitem{CoeyHfO2}J.M.D.  Coey, M. Venkatesan,P. Stamenov, C. B. Fitzgerald, L. S. Dorneles Phys. Rev. B{\bf  72}, 024450 (2005).
\bibitem{Ohno}H. Ohno, Science {\bf 281},951 (1998).
\bibitem{Stoneham} A.M. Stoneham, A.P. Pathak and R.H. Bartram, J. Phys. C: Solid State Phys. {\bf 9}, 73 (1976).
\bibitem{Sawatzky}  I. S. Elfimov, S. Yunoki, and G. A. Sawatzky, Phys. Rev. Lett. {\bf 89}, 216403 (2002).
\bibitem{Sanvito}C. Das Pemmaraju and S. Sanvito, Phys. Rev. Lett. {\bf 94}, 217205 (2005).
\bibitem{Monnier}R. Monnier and B. Delley Phys. Rev. Lett. {\bf 87}, 157204 (2001).
\bibitem{Zunger}J. Osorio-Guillen, S. Lany, S.V. Barabash and A. Zunger, Phys. Rev. Lett. {\bf 96}, 107203 (2006)(We thank a referee for this recent reference).
\bibitem{Vanderbilt}  X. Zhao and D. Vanderbilt, cond-mat/0301016.
\bibitem{Martin}A. K. McMahan, J.F. Annett, R.M. Martin,  Phys. Rev. B {\bf 42}, 6268 (1990).
\bibitem{BouzerarCurrents}G. Bouzerar and D. Poilblanc, Phys. Rev. B {\bf 52}, 10772 (1995).
\bibitem{Grabert}B. Reusch and H. Grabert, Phys. Rev. B {\bf 68}, 045309 (2003).
\bibitem{Bouzerar2}  G. Bouzerar, T. Ziman and J. Kudrnovsk\`y, Europhys. Lett., {\bf 69}, 812-818 (2005); G. Bouzerar, T. Ziman and J. Kudrnovsk\`y, Appl. Physics Lett. {\bf 85} 4941 (2004); G. Bouzerar, T. Ziman and J. Kudrnovsk\`y, Phys. Rev. B {\bf  72}, 125207 (2005).  
\bibitem{Lichtenstein}A.I. Lichtenstein, M.I. Katsnelson and V.A. Gubanov, J. Phys.F. {\bf 14}, L125 (1984),M.I. Katsnelson and A.I. Lichtenstein, Phys. Rev. B {\bf 61}, 8906 (2000).
\bibitem{RichardBouzerar}  R. Bouzerar, G. Bouzerar, T. Ziman 
 Phys. Rev. B {\bf 73}, 024411 (2006)
\bibitem{Josef}J. Kudrnovsk\'y, I. Turek, V. Drchal, F. Maca, P. Weinberger and P. Bruno  Phys. Rev. B {\bf 69}, 115208 (2004).
\bibitem{Kanamori} J. Kanamori, J. Phys. Chem. Solids {\bf 10}, 87 (1959). 
\bibitem{Tasaki}H. Tasaki, Phys. Rev. Lett. {\bf 75}, 4678 (1995).
\bibitem{Nagaoka}  Y. Nagaoka, Phys. Rev. {\bf 147}, 392 (1966). 
\bibitem{Mielke}   A. Mielke, Phys. Rev. Lett. 82, 4312 (1999)
\bibitem{Ohkawa}F. J. Ohkawa Phys. Rev. B {\bf 65}, 174424 (2002).
\end{thebibliography}
\end{document}